\documentclass{fundam}
\usepackage{graphicx}

\begin{document}
\issue{{\large \sc ** In Preparation **}}

\title{A New Algorithm for Building Alphabetic Minimax Trees}
\address{Dipartimento di Informatica,
    via Bellini 25g,
    15100 Alessandria (AL),
    Italy}
\author{Travis Gagie\thanks
    {Supported by Italy-Israel FIRB grant ``Pattern Discovery Algorithms in Discrete Structures, with Applications to Bioinformatics''.}\\
    Department of Computer Science\\
    University of Eastern Piedmont\\
    15100 Alessandria (AL), Italy\\
    travis{@}mfn.unipmn.it}
\runninghead{T. Gagie}{Building Alphabetic Minimax Trees}
\maketitle

\begin{abstract}
We show how to build an alphabetic minimax tree for a sequence \(W = w_1, \ldots, w_n\) of real weights in \(O (n d \log \log n)\) time, where $d$ is the number of distinct integers \(\lceil w_i \rceil\).  We apply this algorithm to building an alphabetic prefix code given a sample.
\end{abstract}


\section{Introduction} \label{sec:introduction}

For the alphabetic minimax tree problem, we are given a sequence \(W = w_1, \ldots, w_n\) of weights and an integer \(t \geq 2\) and asked to find an ordered $t$-ary tree on $n$ leaves such that, if the depths of the leaves from left to right are \(\ell_1, \ldots, \ell_n\), then \(\max_{1 \leq i \leq n} \{w_i + \ell_i\}\) is minimized.  Such a tree is called a $t$-ary alphabetic minimax tree for $W$ and the minimum maximum sum, \(\alpha (W)\), is called the $t$-ary alphabetic minimax cost of $W$.

Hu, Kleitman and Tamaki~\cite{HKT79} gave an \(O (n \log n)\)-time algorithm for this problem when $t$ is 2 or 3.  Under the assumption the tree must be strictly $t$-ary, Kirkpatrick and Klawe~\cite{KK85} gave \(O (n)\)-time and \(O (n \log n)\)-time algorithms for integer and real weights, respectively, which they applied to bounding circuit fan-out.  Coppersmith, Klawe and Pippenger~\cite{CKP86} modified Kirkpatrick and Klawe's algorithms to work without the assumption, and again applied them to bounding circuit fan-out.  Kirkpatrick and Przytycka~\cite{KP90} gave an \(O (\log n)\)-time, \(O (n / \log n)\)-processor algorithm for integer weights in the CREW PRAM model.  Finally, Evans and Kirkpatrick~\cite{EK04} gave an \(O (n)\)-time algorithm for the problem with integer weights in which we want to find a binary tree that minimizes the maximum over $i$ of the sum of the $i$th weight and the $i$th node's (rather than leaf's) depth, and applied it to restructuring ordered binary trees.  In this paper, we give an \(O (n d \log \log n)\)-time algorithm for the original problem with real weights, where $d$ is the number of distinct integers \(\lceil w_i \rceil\).  Our algorithm can be adapted to work for any $t$ but, to simplify the presentation, we assume \(t = 2\) and write $\log$ to mean $\log_2$.

\section{Motivation} \label{sec:motivation}

Our interest in alphabetic minimax trees stems from a problem concerning alphabetic prefix codes, i.e., prefix codes in which the lexicographic order of the codewords is the same as that of the characters.  Suppose we want to build an alphabetic prefix code with which to compress a file (or, equivalently, a leaf-oriented binary search tree with which to sort it), but we are given only a sample of its characters.  Let \(P = p_1, \ldots, p_n\) be the normalized distribution of characters in the file, let \(Q = q_1, \ldots, q_n\) be the normalized distribution of characters in the sample and suppose our codewords are \(C = c_1, \ldots, c_n\).  An ideal code for $Q$ assigns the $i$th character a codeword of length \(\log (1 / q_i)\) (which may not be an integer), and the average codeword's length using such a code is \(H (P) + D (P \| Q)\), where \(H (P) = \sum_i p_i \log (1 / p_i)\) is the entropy of $P$ and \(D (P \| Q) = \sum_i p_i \log (p_i / q_i)\) is the relative entropy between $P$ and $Q$.

Consider the best worst-case bound we can achieve on how much the average codeword's length exceeds \(H (P) + D (P \| Q)\).  As long as \(q_i > 0\) whenever \(p_i > 0\), the average codeword's length is
\begin{eqnarray*}
\sum_i p_i |c_i|
& = & \sum_i p_i \left( \rule{0ex}{2ex}
    \log (1 / p_i) + \log (p_i / q_i) + \log q_i + |c_i| \right)\\
& = & H (P) + D (P \| Q) + \sum_i p_i (\log q_i + |c_i|)
\end{eqnarray*}
(if \(q_i = 0\) but \(p_i > 0\) for some $i$, then our formula is undefined).  Notice each $|c_i|$ is the length of the $i$th branch in the trie for $C$.  Therefore, the best bound we can achieve is
\begin{eqnarray*}
\lefteqn{\min_C \max_P \left\{ \sum_i p_i (\log q_i + |c_i|) \right\}}\\
& = & \min_C \max_i \{\log q_i + |c_i|\}\\
& = & \alpha (\log q_1, \ldots, \log q_n)\,,
\end{eqnarray*}
and we achieve it when the trie for $C$ is an alphabetic minimax tree for \(\log q_1, \ldots, \log q_n\).

In several reasonable special cases, we can build the alphabetic minimax tree for \(\log q_1, \ldots, \log q_n\) in \(o (n \log n)\) time.  For example, if each pair $q_i$ and $q_j$ differ by at most a multiplicative constant --- a case Klawe and Mumey~\cite{KM95} considered when building optimal alphabetic prefix codes --- then each pair \(\log q_i\) and \(\log q_j\) differ by at most an additive constant, so the number of distinct integers \(\lceil \log q_i \rceil\) is constant and our algorithm runs in \(O (n \log \log n)\) time.

\section{Algorithm} \label{sec:algorithm}

Let \(B = b_1, \ldots, b_n\) be the values \(w_1 - \lfloor w_1 \rfloor, \ldots, w_n - \lfloor w_n \rfloor\) sorted into nondecreasing order.  Kirkpatrick and Klawe showed that, if $i$ is the smallest index such that
\[\alpha \left( \rule{0ex}{2ex} \lceil w_1 - b_i \rceil, \ldots, \lceil w_n - b_i \rceil \right)
= \alpha \left( \rule{0ex}{2ex} \lceil w_1 - b_n \rceil, \ldots, \lceil w_n - b_n \rceil \right) \,,\]
then \(\alpha (W) = \left( \rule{0ex}{2ex} \lceil w_1 - b_i \rceil, \ldots, \lceil w_n - b_i \rceil \right) + b_i\) and any alphabetic minimax tree for \(\lceil w_1 - b_i \rceil, \ldots, \lceil w_n - b_i \rceil\) is an alphabetic minimax tree for $W$.  Their \(O (n \log n)\)-time algorithm for real weights is a simple combination of this fact, binary search and their \(O (n)\)-time algorithm for integer weights: they compute and sort \(w_1 - \lfloor w_1 \rfloor, \ldots, w_n - \lfloor w_n \rfloor\) to obtain $B$, compute an alphabetic minimax tree for the sequence \(\lceil w_1 - b_n \rceil, \ldots, \lceil w_n - b_n \rceil\) of integer weights, and use binary search to find $b_i$; for each step of the binary search, if the candidate value to be tested is $b_j$, then they build an alphabetic minimax tree for the sequence \(\lceil w_1 - b_j \rceil, \ldots, \lceil w_n - b_j \rceil\) of integer weights and compare \(\alpha \left( \rule{0ex}{2ex} \lceil w_1 - b_j \rceil, \ldots, \lceil w_n - b_j \rceil \right)\) to \(\alpha \left( \rule{0ex}{2ex} \lceil w_1 - b_n \rceil, \ldots, \lceil w_n - b_n \rceil \right)\).

Our idea is to avoid sorting \(w_1 - \lfloor w_1 \rfloor, \ldots, w_n - \lfloor w_n \rfloor\) and then building an alphabetic minimax tree from scratch for each step of the binary search.  To avoid sorting, we use a technique similar to the one Klawe and Mumey described for generalized selection; to avoid building the trees from scratch, we use a data structure based on Kirkpatrick and Przytycka's level tree data structure for $W$.  Our data structure, which we describe in Section~\ref{sec:data structure}, stores $W$ and \(X = x_1, \ldots, x_n = 0, \ldots, 0\) and performs any sequence of \(O (n)\) of the following operations in \(O (n d \log \log n)\) time:
\begin{description}
\item[{\sf set}$(i)$] --- set $x_i$ to 1;
\item[{\sf undo}] --- undo the last {\sf set} operation;
\item[{\sf cost}] --- return \(\alpha \left( \rule{0ex}{2ex} \lceil w_1 \rceil - x_1, \ldots, \lceil w_n \rceil - x_n \right)\).
\end{description}

We first find \(b_n = \max_i \{w_i - \lfloor w_i \rfloor\}\) and then, using Kirkpatrick and Klawe's \(O (n)\)-time algorithm, \(\alpha \left( \rule{0ex}{2ex} \lceil w_1 - b_n \rceil, \ldots, \lceil w_n - b_n \rceil \right)\).  We build the multiset \(S_0 = \left\{ \rule{0ex}{2ex} \langle w_i - \lfloor w_i \rfloor, i \rangle \right\}\) and use binary search to find the smallest value \(w_i - \lfloor w_i \rfloor\) such that
\begin{eqnarray*}
\lefteqn{\alpha \left( \rule{0ex}{2ex} \lceil w_1 - (w_i - \lfloor w_i \rfloor) \rceil, \ldots,
    \lceil w_n - (w_i - \lfloor w_i \rfloor) \rceil \right)}\\
& = & \alpha \left( \rule{0ex}{2ex} \lceil w_1 - b_n \rceil, \ldots, \lceil w_n - b_n \rceil \right) \,.
\end{eqnarray*}
Once we have \(w_i - \lfloor w_i \rfloor\), we use Kirkpatrick and Klawe's \(O (n)\)-time algorithm again to build an alphabetic minimax tree for the sequence \(\lceil w_1 - (w_i - \lfloor w_i \rfloor) \rceil, \ldots, \lceil w_n - (w_i - \lfloor w_i \rfloor) \rceil\) of integer weights.

For the $k$th step of the binary search, we use Blum et al.'s algorithm~\cite{BFP+73} to find the median $m_k$ of the first components in $S_k$; we divide $S_k$ into
\begin{eqnarray*}
S_k' & = & \left\{ \rule{0ex}{2ex} \langle w_i - \lfloor w_i \rfloor, i \rangle\,:\,
    w_i - \lfloor w_i \rfloor < m_k \right\}\,,\\
S_k'' & = & \left\{ \rule{0ex}{2ex} \langle w_i - \lfloor w_i \rfloor, i \rangle\,:\,
    w_i - \lfloor w_i \rfloor = m_k \right\}\,,\\
S_k'''& = & \left\{ \rule{0ex}{2ex} \langle w_i - \lfloor w_i \rfloor, i\rangle\,:\,
    w_i - \lfloor w_i \rfloor > m_k \right\}\,;
\end{eqnarray*}
for each second component $j$ in $S_k'$ or $S_k''$ with $w_j$ not an integer, we set $x_j$ to 1; we compare \(\alpha \left( \rule{0ex}{2ex} \lceil w_1 \rceil - x_1, \right.\)
\(\left. \rule{0ex}{2ex} \ldots, \lceil w_n \rceil - x_n \right)\) to \(\alpha \left( \rule{0ex}{2ex} \lceil w_1 - b_n \rceil, \ldots, \lceil w_n - b_n \rceil \right)\); if it is equal, then $m_k$ is still a candidate, so we undo all the {\sf set} operations we performed in this step and recurse on $S_k'$; if it is greater, then $m_k$ is too small, so we leave all the {\sf set} operations and recurse on $S_k'''$.  The last candidate considered during the search is the value \(w_i - \lfloor w_i \rfloor\) we want.  For the $k$th step of the search, we spend \(O (n / 2^k)\) time finding the median $m_k$ and dividing $S_k$ into $S_k'$, $S_k''$ and $S_k'''$, and perform \(O (n / 2^{k})\) operations on the data structure.  Summing over the steps, we use \(O (n)\) time to find all the medians and divide all the sets and \(O (n d \log \log n)\) time to perform all the operations on the data structure.

\begin{lemma} \label{lem:algorithm}
Given a data structure that performs any sequence of \(O (n)\) {\sf set}, {\sf undo} and {\sf cost} operations in \(O (n d \log \log n)\) time, we can build an alphabetic minimax tree for $W$ in \(O (n d \log \log n)\) time.
\end{lemma}

\section{Data structure} \label{sec:data structure}

If we define the weight of the $i$th leaf of an alphabetic minimax tree for $W$ to be $w_i$, and the weight of each internal node to be the maximum of its children's weights plus 1, then the weight of the root is \(\alpha (W)\).  We would like to use this property to recompute \(\alpha \left( \rule{0ex}{2ex} \lceil w_1 \rceil - x_1, \ldots, \lceil w_n \rceil - x_n \right)\) efficiently after updating $X$, but even small changes can greatly affect the shape of the alphabetic minimax tree: e.g., suppose \(n = 2^k + 1\), each \(w_i = k - 1 / 2\) and each \(x_i = 0\); if we set $x_1$ and $x_2$ to 1 then, in the unique alphabetic minimax tree for
\[\lceil w_1 \rceil - x_1, \ldots, \lceil w_n \rceil - x_n
= k - 1, k - 1, k, \ldots, k\,,\]
every even-numbered leaf except the second is a left-child; but if we instead set $x_{n - 1}$ and $x_n$ to 1 then, in the unique alphabetic minimax tree for
\[\lceil w_1 \rceil - x_1, \ldots, \lceil w_n \rceil - x_n
= k, \ldots, k, k - 1, k - 1\,,\]
every even-numbered leaf except the \((n - 1)\)st is a right-child.

\begin{figure}
\begin{centering}
\includegraphics[width=40ex]{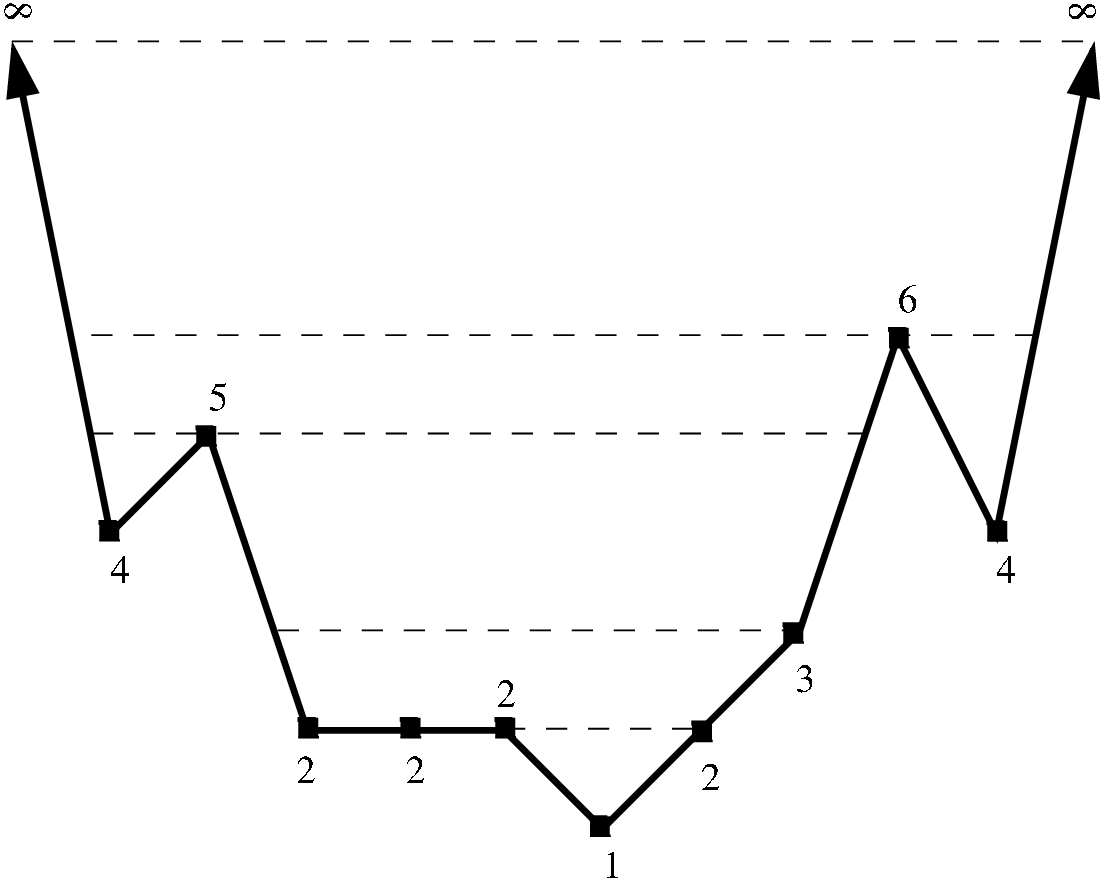}
\caption{The level intervals for \(4, 5, 2, 2, 2, 1, 2, 3, 6, 4\).}
\label{fig:intervals}
\vspace{5ex}
\includegraphics[width=40ex]{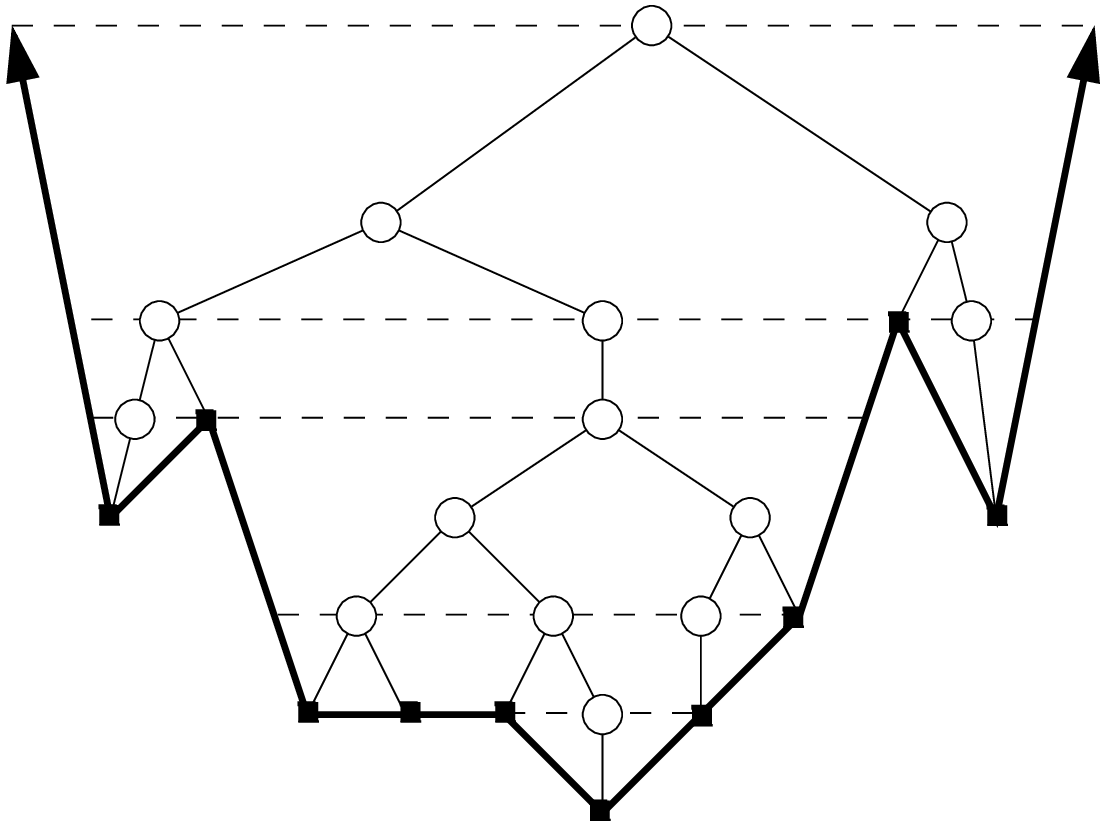}
\caption{An alphabetic minimax tree for \(4, 5, 2, 2, 2, 1, 2, 3, 6, 4\).}
\label{fig:minimax tree}
\vspace{5ex}
\includegraphics[width=40ex]{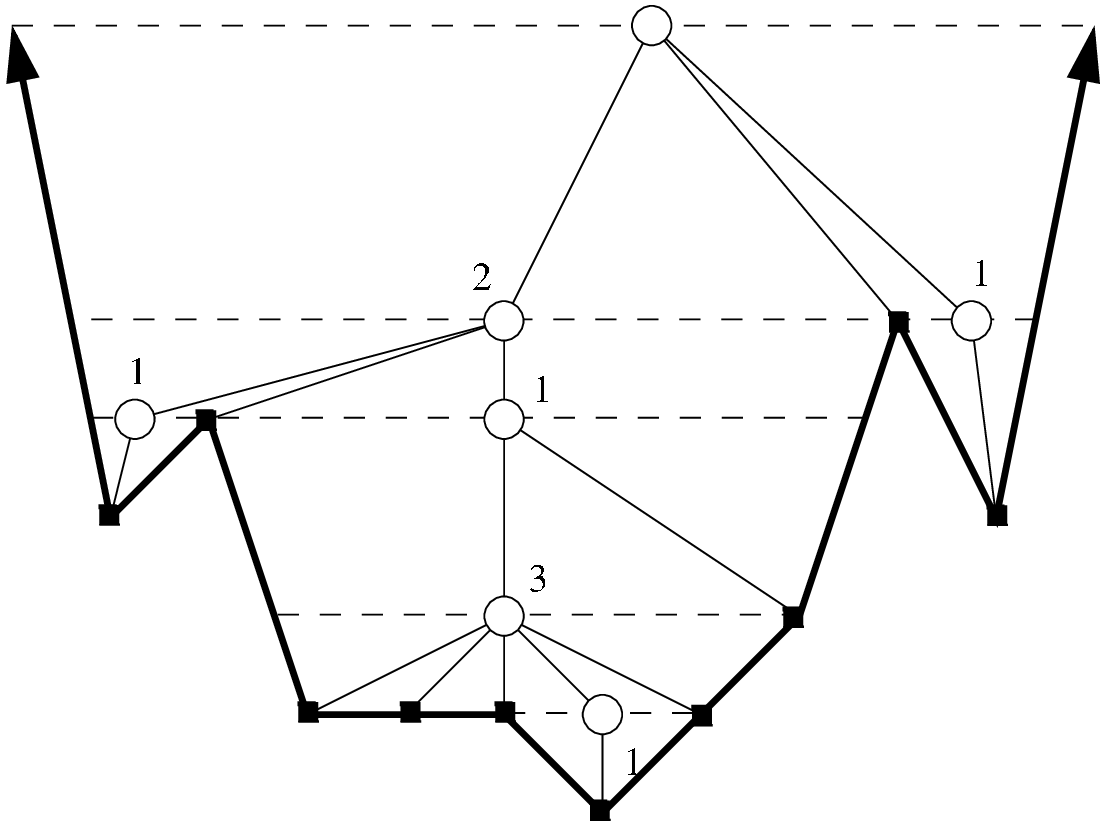}
\caption{The level tree for \(4, 5, 2, 2, 2, 1, 2, 3, 6, 4\), with internal nodes' loads shown.}
\label{fig:level tree}
\end{centering}
\end{figure}

Fortunately for us, Kirkpatrick and Przytycka defined a data structure, called a level tree, that represents an alphabetic minimax tree but whose shape is less volatile.  Let
\[Y
= y_1, \ldots, y_n
= \lceil w_1 \rceil - x_1, \ldots, \lceil w_n \rceil - x_n\,,\]
and consider their definition of the level tree for $Y$ (we have changed their notation slightly to match our own):
\begin{quotation}
``We start our description of the level tree with the following geometric construction (see Figure~\ref{fig:intervals}): Represent the sequence of weights $Y$ by a polygonal line; for every \(i = 1, \ldots, n\) draw on the plane the point \((i, y_i)\), and for every \(i = 1, \ldots, n - 1\) connect the points \((i, y_i)\) and \((i + 1, y_{i + 1})\); for every $i$ such that \(y_i > y_{i + 1}\) (resp., \(y_i > y_{i - 1}\)) draw a horizontal line going from \((i, y_i)\) to its right (resp., left) until it hits the polygonal line.  The intervals defined in such a way are called the \emph{level intervals}.  We also consider the interval \([(0, \infty), (n + 1, \infty)]\) and the degenerate intervals \([(i, y_i), (i, y_i)]\) as level intervals.  Let $e$ be a level interval.  Note that at least one of $e$'s endpoints is equal to \((i, y_i)\) for some index $i$. \dots We define the \emph{level of a level interval} to be equal to [the second component of points belonging to that interval].

Note that an alphabetic minimax tree can be embedded in the plane in such a way that the root of the tree belongs to the level interval \([(0, \infty), (n + 1, \infty)]\) and that internal nodes whose weights are equal to the weight of one of the leaves belong to the horizontal line through this leaf.  Furthermore, if there is a tree edge cutting a level interval then adding a node subdividing this edge to the alphabetic minimax tree does not increase the weight of the root.  By this observation we can consider alphabetic minimax trees which can be embedded in the plane in such a way that all edges intersect level intervals only at endpoints (see Figure~\ref{fig:minimax tree}).

The \emph{level tree} for $Y$ is the ordered tree whose nodes are in one-to-one correspondence with the level intervals defined above.  The parent of a node $v$ is the internal node which corresponds to the closest level interval which lies above the level interval corresponding to $v$.  The left-to-right order of the children of an internal node corresponds to the left-to-right order of the corresponding level intervals on the plane (see Figure~\ref{fig:level tree}).  For every node $u$ of a level tree we define \(\mathrm{load} (u)\) to be equal to the number of nodes of the constructed alphabetic minimax tree which belong to the level interval corresponding to $u$ (assuming the above embedding).

If $u$ is a leaf then \(\mathrm{load} (u) = 1\).  Assume that $u$ is an internal node and let \(u_1, \ldots, u_k\) be the children of $u$.  Let \(\Delta_u\) denote the minimum of the value \(\lceil \log n \rceil\) and the difference between the level of the level interval corresponding to node $u$ and the level of the intervals corresponding to its children.  It is easy to confirm that
\[\mathrm{load} (u)
= \left\lceil \frac{\mathrm{load} (u_1) + \cdots + \mathrm{load} (u_k)}
    {2^{\Delta_u}} \right\rceil\,.\mbox{''}\]
\end{quotation}
Notice that, if $u$ is the root of the level tree and \(u_1, \ldots, u_k\) are its children, then Kirkpatrick and Przytycka embed \(\mathrm{load} (u_1) + \cdots + \mathrm{load} (u_k)\) nodes of the alphabetic minimax tree into the intervals corresponding to \(u_1, \ldots, u_k\).  It follows that \(\alpha (Y)\) is the level of the intervals corresponding to \(u_1, \ldots, u_k\) plus \(\left\lceil \rule{0ex}{2ex} \log (\mathrm{load} (u_1) + \cdots + \mathrm{load} (u_k)) \right\rceil\).

It is straightforward to build the level tree for $Y$ in \(O (n)\) time, by first building an alphabetic minimax tree for it.  Moreover, if we set a bit $x_i$ to 1 and thus decrement $y_i$, then the shape of the level tree for $Y$ and the loads change only in the vicinity of the $i$th leaf and along the path from it to the root.  The number of levels is the number of distinct weights in $Y$ plus one, so the length of that path is \(O (d)\) (recall $d$ is the number of distinct integers $\lceil w_i \rceil$).  Unfortunately, the level tree can have very high degree, so we may not be able, e.g., to navigate very quickly from the root to a leaf.

\begin{figure}
\begin{centering}
\includegraphics[width=50ex]{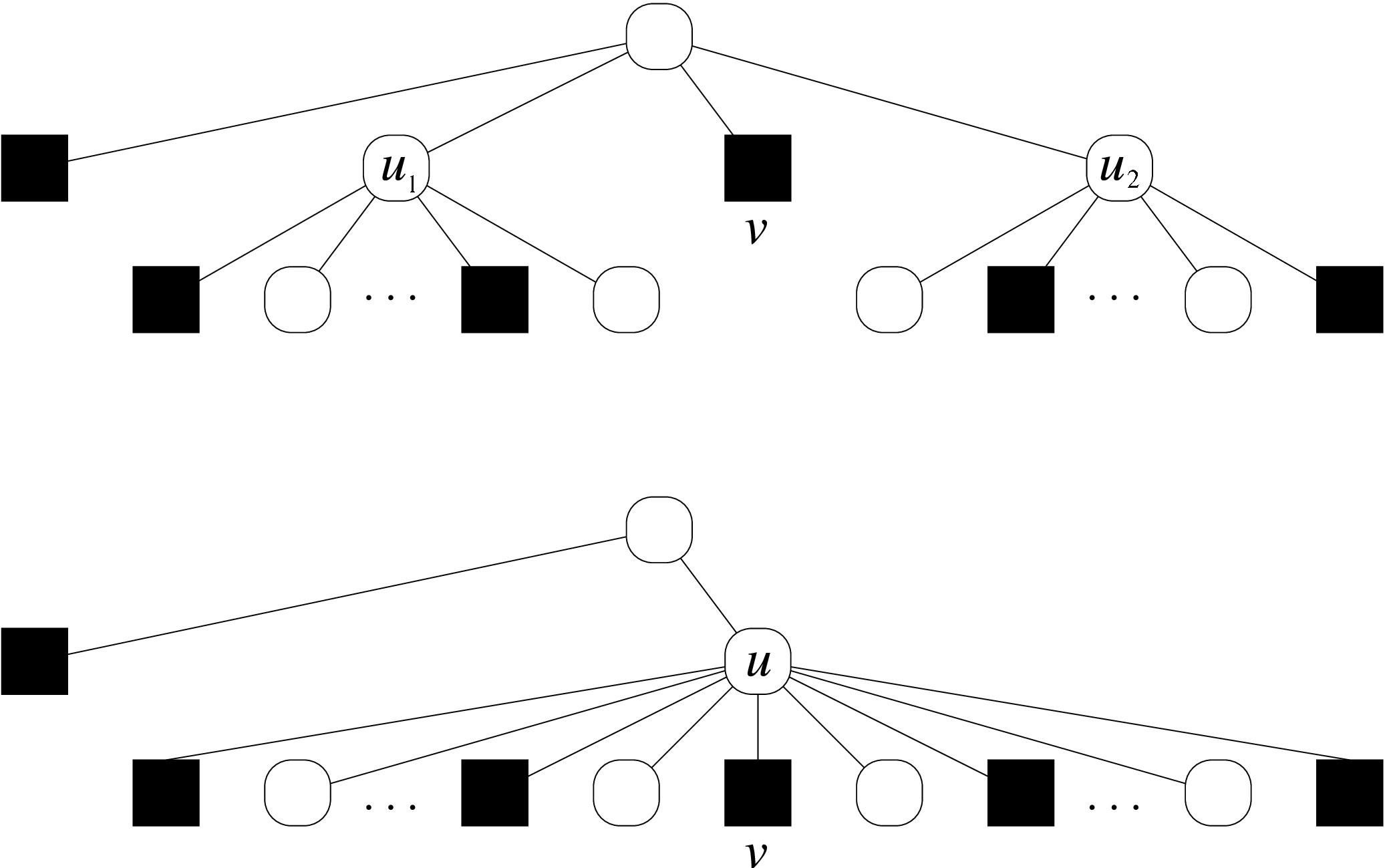}
\caption{Decrementing a node $v$'s level can force us to combine its adjacent siblings $u_1$ and $u_2$ into a new node $u$.}
\label{fig:set}
\end{centering}
\end{figure}

We store a pointer to the root of the level tree and an array of pointers to its leaves, and pointers from each node to its parent.  At each internal node, we store its children in a doubly-linked list (so each child points to the siblings immediately to its left and right).  It is not hard to verify that, with these pointers, we can implement a {\sf cost} operation in \(O (1)\) time and reach all the nodes that need to be updated for a {\sf set} operation in \(O (d)\) time.  We cannot implement {\sf set} operations in \(O (d)\) worst-case time, however, because of the following case (see Figure~\ref{fig:set}): suppose the siblings $u_1$ and $u_2$ immediately to the left and right of the $i$th leaf $v$ are internal nodes whose children belong to level intervals with level \(y_i - 1\); if we set $x_i$ to 1 and thus decrement $y_i$ and $v$'s level, then $u_1$'s former children, $v$ and $u_2$'s former children will all have the same parent (either a new node $u$ if $v$ had siblings other than $u_1$ and $u_2$, as shown in Figure~\ref{fig:set}, or their former parent if it did not).

To deal with this case, we store all the internal nodes of the level tree in a union-find data structure, due to Mannila and Ukkonen~\cite{MU86}, that supports a {\sf deunion} operation.  Rather than adjusting all of $u_1$'s and $u_2$'s former children to point to their new parent, we simply perform a {\sf union} operation on $u_1$ and $u_2$.  Whenever we follow a pointer to an internal node, we perform a {\sf find} operation on it and, if necessary, update the pointer.  Each {\sf cost} operation on the level tree takes one {\sf find} operation on the union-find data structure and \(O (1)\) extra time, and each {\sf set} operation takes at most one {\sf union} operation, \(O (d)\) {\sf find} operations and \(O (d)\) extra time.  Whenever we make a modification to the level tree other than an operation on the union-find data structure, we push it onto a stack.  To perform an {\sf undo} operation on the level tree, we pop and reverse all the modifications we made since starting the last {\sf set} operation and, if necessary, perform a {\sf deunion} operation.  Any sequence of \(O (n)\) operations on the level tree takes \(O (n d)\) operations on the union-find data structure, which Mannila and Ukkonen showed take a total of \(O (n d \log \log n)\) time.

\begin{lemma} \label{lem:data structure}
In \(O (n)\) time we can build a data structure that performs any sequence of \(O (n)\) {\sf set}, {\sf undo} and {\sf cost} operations in \(O (n d \log \log n)\) time.
\end{lemma}

\section{Conclusion} \label{sec:conclusion}

Combining Lemmas~\ref{lem:algorithm} and~\ref{lem:data structure}, we have the following theorem:
\begin{theorem} \label{thm:conclusion}
We can build an alphabetic minimax tree for $W$ in \(O (n d \log \log n)\) time.
\end{theorem}
Since $d$ could be as small as 1 or as large as $n$, our theorem is incomparable to previous results.  We can build the tree in \(O \left( \rule{0ex}{2ex} n \min (d \log \log n, \log n) \right)\) time, of course, by first finding $d$ in \(O (n)\) time and then, depending on whether \(d \log \log n < \log n\), using either our algorithm or one of the \(O (n \log n)\)-time algorithms mentioned in Section~\ref{sec:introduction}.

In closing, we note there has recently been interesting work involving unordered minimax trees.  Baer~\cite{Bae08} observed that the problem of building a prefix code with mimimum maximum pointwise redundancy --- originally posed and solved by Drmota and Szpankowski~\cite{DS04} --- can also be solved with a Huffman-like algorithm, due to Golumbic~\cite{Gol76}, for building unordered minimax trees.  Given a probability distribution over $n$ characters, Drmota and Szpankowski's algorithm takes \(O (n \log n)\) time, or \(O (n)\) time if the probabilities are sorted by the fractional parts of their logarithms; we conjecture that, by using Blum et al.'s algorithm as we did in this paper, it can be made to run in \(O (n)\) time even when the probabilities are unsorted.  Like Huffman's algorithm (see~\cite{Van76}), Golumbic's algorithm takes \(O (n \log n)\) time, or \(O (n)\) time if the probabilities are sorted by their values.

\bibliographystyle{fundam}
\bibliography{minimax}

\end{document}